\documentclass[aps,prl,showpacs,twocolumn,groupedaddress]{revtex4}
\usepackage{graphicx,amsmath,verbatim}

\DeclareMathOperator\sgn{sign}

\begin{document}

\title{Quantum Hall ferromagnetic states and spin-orbit interactions in the fractional regime}
\author{Stefano Chesi}
\author{Daniel Loss}
\affiliation{Department of Physics, University of Basel, 
CH-4056 Basel, Switzerland}

\date{\today}

\begin{abstract}
The competition between the Zeeman energy and the Rashba and Dresselhaus spin-orbit couplings is studied for fractional quantum Hall states by including correlation effects. A transition of the direction of the spin-polarization is predicted at specific values of the Zeeman energy. We show that these values can be expressed in terms of the pair-correlation function, and thus provide information about the microscopic ground state. We examine the particular examples of the Laughlin wavefunctions and the 5/2-Pfaffian state. We also include effects of the nuclear bath.
\end{abstract}


\maketitle

\setlength\arraycolsep{0pt}

Two-dimensional electrons in strong magnetic fields have been a rich source of new physics, a prominent example being the discovery of fractional quantum Hall states \cite{JainCambridge07,TheBook}. At large cyclotron energy the ground state is well approximated assuming a small number of completely filled low Landau levels (LLs) while the large degeneracy of the partially filled highest LL is resolved by the electron interaction. 

Additional spin degeneracy is obtained at vanishing Zeeman coupling, realized in GaAs/AlGaAs heterostructures by confinement \cite{snelling91}, hydrostatic pressure \cite{maude96}, or gate modulation \cite{salis01}. Under these conditions, the ground state can still be spin polarized due to the Coulomb interaction, but the polarization \emph{direction} is determined by small spin anisotropies induced by the spin-orbit interaction. The effect of the spin-orbit coupling in the quantum Hall regime was studied in \cite{falko00,schliemann03,desrat05,zarea05,sherman06,valinrodriguez06,lipparini06}. There, it was shown that below a critical value of the Zeeman energy the spin polarization deviates from the perpendicular direction and acquires an in-plane component. The previous treatment, however, was restricted to the case of integer filling factors while we examine here the fractional regime. This represents a nontrivial extension due to the highly correlated nature of the fractional wavefunctions, as opposed to the integer quantum Hall states. Furthermore, we obtain the effect of the simultaneous presence of Rashba and Dresselhaus spin-orbit couplings \cite{bychkov84b,dresselhaus55}.

As a main result, we find that by including correlation effects the polarization transition explicitly depends on the quantum Hall ground state, and to leading order is determined by the pair-correlation function. This provides a new way to address many-body properties of the wavefunctions in the fractional regime. In fact, polarization measurements can be performed with established experimental techniques, as in particular photoluminescence \cite{kukushkin99} or NMR studies \cite{melinte00}. Furthermore, polarization properties are generally less affected by disorder \cite{JainCambridge07} (in contrast to e.g. gap measurements).     

Our discussion is generally applicable to polarized quantum Hall states. We consider here the Laughlin wavefunctions and the Pfaffian state at $\nu=5/2$ \cite{willett87, pan99,morf98,rezayi00}. The latter has received special attention \cite{miller07, nayak07, wan07, dolev08,radu08} since it might support excitations with non-abelian  statistics \cite{moore91}. This proposal is consistent with the recent observation of $e/4$ charged quasiparticles \cite{dolev08,radu08}.

Let us assume a high-field ground state with a partially occupied highest LL which is fully spin polarized along an arbitrary direction $\vec n$. Further, a certain number $J$ of lower LLs are fully occupied for both spin orientations. The anisotropy in the polarization direction $\vec n$ is determined by the Zeeman energy and a general combination of Rashba ($\alpha$) and Dresselhaus ($\beta$) spin-orbit interactions 
\begin{equation}\label{Hpert}
\delta\hat H=\alpha (\hat \pi_x \hat \sigma_y-\hat \pi_y \hat \sigma_x)+
\beta (\hat \pi_x \hat \sigma_x-\hat \pi_y \hat \sigma_y)
-\frac{g \mu_B B}{2}\hat \sigma_z ~,
\end{equation}
where $B>0$ (an opposite polarization is obtained if the magnetic field $\vec{B}$ is along $+\hat z$), and $\hat{\mbox{\boldmath $\pi$}}/m$ is the standard kinematic velocity \cite{JainCambridge07, TheBook}. Second-order perturbation theory in the spin-orbit interaction gives us the angular-dependent energy correction, expressed in terms of the spherical coordinates $(\theta,~\varphi)$ of $\vec n$ 
\begin{eqnarray}\label{energy_correction}
\frac{\delta E}{p N}= &&\,
\left\{-\frac{g \mu_B B}{2}+[2 J+1-\eta f_1(\eta,\nu)]
m (\alpha^2-\beta^2)\right\}\cos\theta \nonumber \\
&&-\frac12 \, \eta f_2(\eta,\nu) \,m (\alpha^2+\beta^2+ 
2 \alpha \beta \sin 2\varphi)\sin^2\theta  ~.
\end{eqnarray}
In Eq.~(\ref{energy_correction}), $N$ is the total number of electrons, $p=(\nu-2 J)/\nu$ is the polarization without spin-orbit coupling, and we defined the interaction parameter \cite{comment_Coulomb} $\eta=(e^2/\epsilon \ell)/\hbar\omega_c$, where $\omega_c=e B/m c$, $\ell=\sqrt{\hbar c/e B}$ is the magnetic length, and $\epsilon$ the dielectric constant. The expressions for $f_{1,2}$ are provided in \cite{EPAPSmy} and we discuss later their explicit form to leading order in $\eta$. 

\begin{figure}
\includegraphics[width=0.4\textwidth]{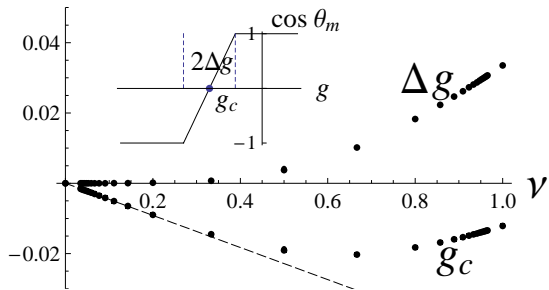}
\caption{\label{gc_vs_nu_lowesLL} Values of $g_c$ (negative dots) and $\Delta g$ (positive dots), as obtained
from Eqs.~(\ref{gc_eq}) and (\ref{dg_eq})
for states at $\nu=1/M$ (and their particle-hole conjugates), and $\nu=1/2$. 
The density is chosen such that $B=5/\nu$ (in T), $\alpha=0$, and other parameters are given in the main text. 
The dashed line is the noninteracting value of $g_c$ ($\Delta g=0$ for noninteracting electrons).
Inset: general plot of the polarization angle $\theta_m$ [see Eq.~(\ref{theta})] as a function of the $g$-factor. 
}
\end{figure}

From Eq.~(\ref{energy_correction}) we immediately obtain that the polarization $\vec n$ satisfies $\sin 2\varphi_m=\sgn(\alpha \beta)$ (we find $f_2>0$). The anisotropy in $\varphi$ disappears for $\alpha=0$ or $\beta=0$ \cite{falko00,schliemann03}. The polarization $\vec n$ can be tilted from the vertical direction, in which case $\theta$ is given by
\begin{equation}\label{theta}
\cos\theta_m=
\frac{g \mu_B B - 2 m \gamma_-^2[2 J+1-\eta f_1(\eta,\nu)]}
{2m \gamma_+^2 \, \eta f_2(\eta,\nu) }~,
\end{equation}
where $\gamma_\pm^2=(|\alpha|+|\beta|)(|\alpha|\pm|\beta|)$. It is easiest to consider the case in which the $g$-factor is changed at fixed external parameters \cite{snelling91, maude96, salis01}. The transition from $-\hat z$ to $+\hat z$ is illustrated in Fig.~\ref{gc_vs_nu_lowesLL}. It occurs around $g_c$, which corresponds to the condition of in-plane polarization $\vec n$
\begin{equation}
\label{gc_eq}
g_c=\frac{2m\gamma_-^2}{\mu_B B}[2J+1-\eta f_1(\eta,\nu)]~,
\end{equation}
and the transition region has a width of $2\Delta g$, where
\begin{equation}
\label{dg_eq}
\Delta g=\frac{2m \gamma_+^2}{\mu_B B} \, \eta f_2(\eta,\nu) ~.
\end{equation}
At $\alpha=\beta=0$ one has as usual $g_c=\Delta g = 0$, while for the non-interacting problem with spin-orbit interaction $g_c\neq 0$ but still $\Delta g=0$. Thus, the Coulomb interaction results in a shift $\sim \eta f_1$ in $g_c$ and opens a finite region $\sim \eta f_2$ in which the polarization $\vec n$ acquires an in-plane component. Note also that from Eqs.~(\ref{gc_eq}, \ref{dg_eq}) we obtain $g_c = 0$, but still $\Delta g \neq 0$, in the special case $|\alpha|=|\beta|$.

The effect of an in-plane component of the external field is straightforward to include in Eq.~(\ref{energy_correction}), via a term $\frac12 g \mu_B B_\parallel \cos(\varphi-\varphi_\parallel)\sin\theta$. This results in a correction to the equilibrium values $(\theta_m,\varphi_m)$ of the polarization $\vec n$ which is anisotropic in the field angle $\varphi_\parallel$. 

A calculation of $f_{1,2}$ is in general difficult, but the leading contribution in $\eta$ can be obtained explicitly in terms of `generalized' pair-correlation functions $\left. g\right._{i,j}^{h,k}({\bf r}_1,{\bf r}_2)$ \cite{EPAPSmy}. These are generated iteratively from the ordinary pair-correlation function, which we parametrize as in \cite{girvin84} 
\begin{equation}\label{pair-corr}
\left.g\right._{0,0}^{0,0}= 
1-e^{-|z|^2/2\ell^2}+{\sum_m}^{\prime}
\frac{2}{m!}\left( \frac{|z|^2}{4\ell^2}\right)^m c_m e^{-|z|^2/4\ell^2}~,
\end{equation}
where $z=z_1-z_2$ ($z_{\alpha}=x_{\alpha}+i y_{\alpha}$) and the prime 
indicates a summation over positive odd values of $m$ only. It is these $\{ c_m \}$ that characterize the specific ground state and also parametrize the final results for $f_{1,2}$.

We find \cite{EPAPSmy}, at $0<\nu<1$,
\begin{equation}\label{f12_LLL}
f_1(0,\nu)=f_2(0,\nu)=\frac{\nu}{2}\sqrt{\frac{\pi}{2}} +
\nu {\sum_m}^\prime \frac{c_m  \Gamma(m-1/2)}{4 m!} ~,
\end{equation}
and at $2<\nu<3$,
\begin{eqnarray}\label{f1}
f_1(0,\nu)=\frac{3\nu-2}{8}\sqrt{\frac{\pi}{2}} +
(\nu-2) {\sum_m}^\prime \frac{c_m  \Gamma(m-5/2)}{256 m!} \nonumber\\
\times 3 (8m-15)(8m-5)~,\qquad
\end{eqnarray}
\begin{eqnarray}\label{f2}
f_2(0,\nu)=\frac{7(\nu-2)}{8}\sqrt{\frac{\pi}{2}}
+(\nu-2){\sum_m}^\prime \frac{c_m \Gamma(m-5/2)}{256 m!}\nonumber\\
\times (105-112m+64m^2) ~,\qquad
\end{eqnarray}
where the summations are restricted to positive odd integers.
The remarkable result $f_1=f_2$ when $0<\nu<1$ is only established 
to lowest order in $\eta$.

As a first application, we consider now the case of the Laughlin trial wavefunctions, which are appropriate for $\nu=1/M$ where $M$ is an odd integer. For simplicity, we adopted the approximation used in \cite{girvin84}. This amounts to set $c_m=-1$ for $m<M$ and $c_m=0$ for $m>M+4$. The three remaining coefficients are determined by exact sum rules \cite{girvin84}. We obtain $f_{1,2}=0.0710,~0.0301$ for $M=3,5$ respectively. These values show small deviations for more accurate parameterizations of the $\{ c_m \}$ coefficients (e.g. using the $\{ c_m \}$ of \cite{macdonald86} gives $f_{1,2}=0.0708,~0.0300$). The same approximation is used at $\nu=2+1/M$ and, by making use of the particle-hole symmetry \cite{EPAPSmy}, the states at $\nu=1-1/M$ and $\nu=3-1/M$ can also be studied.
\begin{table}
\begin{tabular}{ccc||ccc}
     & CFS & Pf &   & CFS & Pf  \\
\hline
$c_1$  & -0.5699 & -0.4205 & $c_9$    & -0.3518 & 0.8761  \\
$c_3$  & 0.4559  & 0.0333  & $c_{11}$ & 0.9403  & -1.406  \\
$c_5$  & -0.0261 & 0.3521  & $c_{13}$ & -0.7151 & 1.170   \\
$c_7$  & -0.1660  & -0.4853& $c_{15}$ & 0.1825  & -0.3703 
\end{tabular}
\caption{\label{cm_Pf_CF} 
Parameters of the pair-correlation function for the Composite Fermi Sea (CFS) and Pfaffian (Pf) wavefunctions.}
\end{table}

We turn now to the Pfaffian (Pf) trial state, which implies half filling of the highest LL ($\nu=1/2,$ 5/2). A closely related compressible state is the polarized composite Fermi sea (CFS). The Pfaffian state is produced by pairing of free composite Fermions (CFs) \cite{read00,willett02}, due to their residual interaction. Therefore, the quantitative properties of these two trial states are very similar \cite{park98}. We list in Table \ref{cm_Pf_CF} their $\{ c_m\}$ parameterizations, which we obtained by fitting the pair-correlation functions of \cite{park98}. At $\nu=1/2$ we have $f^{CFS}_{1,2}=0.20$ and $f^{Pf}_{1,2}=0.22$, while at $\nu=5/2$
\begin{eqnarray}
\label{f1_52}
f^{CFS}_{1}(0,5/2)=1.16~,  \quad f^{Pf}_{1}(0,5/2)=1.01~,\quad \\
\label{f2_52}
f^{CFS}_{2}(0,5/2)=0.49~, \quad f^{Pf}_{2}(0,5/2)=0.45~. \quad 
\end{eqnarray}
Evidently, the values of $f_{1,2}$ reflect the different correlations of these two trial states. We obtain a $\sim 10\%$ relative change in the values of $f_{1,2}$, which is significantly larger than the change in the total energy \cite{park98}. 

The values of $f_{1,2}$ can be accessed through the measurement of $g_c$ and $\Delta g$ [see Eqs.~(\ref{gc_eq}) and (\ref{dg_eq})], which makes them an experimentally relevant characterization of the quantum Hall state. As it is clear from Eqs.~(\ref{f12_LLL})-(\ref{f2}), $f_{1,2}$ provide information about the $c_m$ coefficients, and therefore on the pair-correlation function. In fact, by truncating the series (\ref{f12_LLL})-(\ref{f2}) to the two lowest terms, an estimate of $c_1$ and $c_3$ from the measured values of $f_{1}$ and $f_{2}$ is obtained. This procedure is justified since the $c_m$ prefactors decrease like $m^{-3/2}$. 

For example, using our `exact' values of $f_{1,2}$ in (\ref{f1_52}) and (\ref{f2_52}) one obtains for the Pfaffian $c_{1,3}\simeq -0.43,~0.07$, in reasonable agreement with Table \ref{cm_Pf_CF} and clearly distinct from the CF sea values. Furthermore, to obtain approximate values of $c_{1,3}$ at $\nu= 1/3$ and $1/5$ would test the distinct short-range behavior of the pair-correlation functions ($\sim r^6$ and $r^{10}$ respectively) of the Laughlin wavefunctions. Also in the controversial case $\nu=7/3$ (see e.g. \cite{dambruenil88}), to measure $c_1$ would provide a test of the Laughlin model at this filling factor.
 
\begin{figure}
\includegraphics[width=0.4\textwidth]{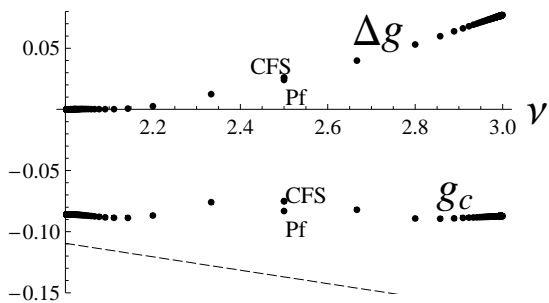}
\caption{\label{gc_vs_nu} Same as Fig.~\ref{gc_vs_nu_lowesLL}, in the range $2<\nu<3$.
At $\nu=5/2$ both Composite Fermi Sea (CFS) and Pfaffian (Pf) results are displayed. 
The density is such that $B=12.5/\nu$ (in T) and $\alpha=0$. The noninteracting $g_c$ is also plotted (dashed line). }
\end{figure}

Let us now estimate the effects for typical GaAs parameters, and thereby demonstrate that our predictions are within experimental reach. We evaluate Eqs.~(\ref{gc_eq}) and (\ref{dg_eq}) using $m=0.067 m_0$, $\epsilon=12.4$, and for a symmetric well with thickness $L=6$ nm, close to the value at which the $g$-factor is zero \cite{snelling91,maude96}. We obtain for the Dresselhaus coupling $\hbar\beta=\lambda (\pi/L)^2 \simeq 27$ meV\AA, where $\lambda\simeq 10$ eV\AA$^3$ \cite{EPAPScubic}, and $\alpha=0$. The results for $g_c$ and $\Delta g$ are plotted in Figs.~\ref{gc_vs_nu_lowesLL} and \ref{gc_vs_nu} in the range $0<\nu<1$ and $2<\nu<3$, respectively. We assumed a constant density $\rho=1.21\times 10^{11}$ cm$^{-2}$ in the first case and $\rho=3.02\times 10^{11}$ cm$^{-2}$ for the second one \cite{willett87}. As seen, the values of $g_c$, $\Delta g$ are in the range already realized in practice \cite{snelling91, maude96, salis01}. 

Figs.~\ref{gc_vs_nu_lowesLL} and \ref{gc_vs_nu} show that the effect of the interaction on the values of $g_c$ can be rather large. This is clearly identified as the deviation of the $g_c$ values (points) from the dashed line, which refers to the noninteracting result. Furthermore, the transition region $\Delta g$ can be sizable, as opposed to the noninteracting case $\Delta g=0$. When $\nu \to 0$ (see Fig.~\ref{gc_vs_nu_lowesLL}), the values of $g_c$ approach the noninteracting linear dependence of Eq.~(\ref{gc_eq}) (with $J=0$, $\eta=0$, and $\gamma_-^2=-\beta^2$). This limit allows one to extract $\beta$, independently from other methods known in the literature. 

We note that the difference between specific realizations at $\nu=5/2$ (CFS and Pf) is small compared to the total effect. Nevertheless, we suggest that the relative change could be detected from temperature-dependent measurements. When the temperature exceeds the pairing energy (but remains still smaller than the CFs kinetic energy), the existence of a CF sea was experimentally demonstrated in \cite{willett02}. We expect that the CFS values of Fig.~\ref{gc_vs_nu} would be observed in this high-temperature regime. With decreasing temperature, $g_c$ and $\Delta g$ evolve into the Pfaffian values, due to the formation of the incompressible state of paired CFs \cite{read00,willett02}. 

\begin{figure}
\includegraphics[width=0.35\textwidth]{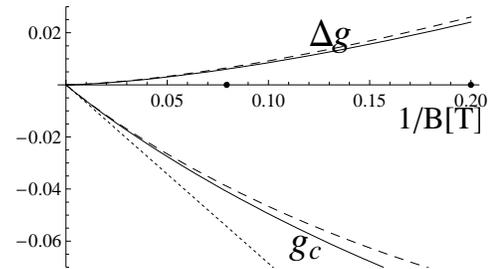}
\caption{\label{gc} Plot of the values of $g_c$ (negative) and $\Delta g$ (positive) 
from Eqs. (\ref{gc_eq}) and (\ref{dg_eq}) at $\nu=5/2$, $B>5~{\rm T}$ and other parameters as in the text. 
The solid and dashed lines correspond to the Pfaffian state and CF sea, resp. [$f_{1,2}$ as in Eqs. 
(\ref{f1_52}, \ref{f2_52})]. The dotted line is the linear noninteracting 
contribution to $g_c$ ($f_1=0$). The dots mark the $B$-field values of \cite{willett87, pan01}.}
\end{figure}

Higher order corrections in $\eta$ affect the precise values of $g_c$ and $\Delta g$. Since $\eta$ is often not particularly small under the typical conditions at which the $\nu=5/2$ state is observed ($\eta \simeq 0.74$ at the highest field $12.6$ T in \cite{pan01}), measurements at larger values of $B$ would be desirable. We show in Fig.~\ref{gc} the high field dependence of $g_c$, $\Delta g$ from Eqs.~(\ref{gc_eq}) and (\ref{dg_eq}). The $f_1$-coefficient determines the $\propto 1/\sqrt{B^3}$ correction to the noninteracting background, which is linear in $1/B$. The $f_2$-coefficient gives the leading $\propto 1/\sqrt{B^3}$ contribution to $\Delta g$.
Alternatively, higher orders in $1/\sqrt{B}$ have to be explicitly computed.

The assumption of full polarization of the highest LL is justified at several values of $\nu$ (e.g. $\nu=1/M$ and $\nu=5/2$). At other fractional values the ground state can be unpolarized (e.g. $\nu=1/2,~2/3$) or partially polarized (e.g. $\nu=3/5,~3/7$). For the latter case, a similar effect is expected, driven to leading order by the noninteracting contribution ($f_{1,2}=0$), but our calculation of $f_{1,2}$ does not apply. It is of conceptual interest to consider a large value of $\beta$, such that full polarization is obtained around $g_c$ in the whole intervals $0<\nu<1$ and $2<\nu<3$. It would then be possible to observe non-analytic features at the incompressible values (see \cite{EPAPSmy}), similarly to the predicted cusps in the total energy \cite{halperin84}.

\begin{figure}
\includegraphics[width=0.4\textwidth]{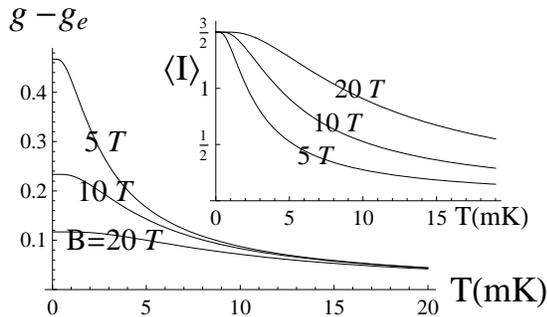}
\caption{\label{deltagN} Nuclear shift of the hyperfined modified electron $g$-factor 
(see text) as a function of temperature $T$ at different values of $B$. 
In the inset, the average nuclear polarization $\langle I \rangle=\sum_i x_i  \langle \hat I_z \rangle_i$
is also shown.}
\end{figure}

Finally, we note that at ultra-low temperatures at which the $\nu = 5/2$ 
state is observed, there is a significant effect from the nuclear spin bath. This contribution can be
easily included by interpreting $g$ in Eq.~(\ref{Hpert}) as $g=g_e-\sum_i x_i A_i \langle \hat I_z \rangle_i/\mu_B B$,
where $g_e$ is the `bare' electron $g$-factor of the heterostructure and the second 
term is the Overhauser shift produced by the hyperfine interaction. Here, $x_i$ are the fractions relative to the different nuclear species (equal to $0.5,~0.3,~0.2$ for ${}^{75}$As, ${}^{69}$Ga, ${}^{71}$Ga, respectively)
and $A_i$ are the corresponding hyperfine couplings (with estimated values~\cite{paget77} $94,~77,~99~\mu$eV).
In Fig. \ref{deltagN} we plot the shift $g-g_e$ as function of $T$ for different values of $B$. 
The high-$T$ limit gives $g-g_e\simeq 0.9/T$ ($T$ in mK), independent of $B$. 
We see that a change of temperature might provide a practical way of tuning the 
small Zeeman energies involved.

We would like to thank R. H. Morf for interesting discussions. Financial support by the NCCR Nanoscience and the Swiss NSF is acknowledged.



\renewcommand{\theequation}{A\arabic{equation}}
\renewcommand{\thefigure}{A\arabic{figure}}
\renewcommand\bibnumfmt[1]{[A#1]}
\newcommand{\citenumfont}{A}

\setcounter{figure}{0}
\setcounter{equation}{0}

\section*{APPENDIX}


We discuss here the general expressions for $f_{1,2}$ and their derivation to leading order in the Coulomb interaction parameter $\eta=(e^2/\epsilon \ell)/\hbar\omega_c$. For a magnetic field along $-\hat z$ we take single-particle wavefunctions of the form
\begin{eqnarray}\label{epaps:psidef1}
\varphi_{0,n}(z)&=&\frac{1}{\sqrt{2\pi \ell^2 n!}}
\left(\frac{z}{\sqrt{2}\ell} \right)^n e^{-|z|^2/4\ell^2}~,\\
\label{epaps:psidef2}
\varphi_{j,n}(z)&=& i\sqrt{\frac{2}{j}} \,
\left(\frac{z^*}{4\ell}-\ell\frac{\partial}{\partial z} \right)\varphi_{j-1,n}(z)~,
\end{eqnarray}
where $z=(x+i y)/\ell$ and $j,n=0,1,2,\ldots$ Here $j$ is the Landau Level (LL) index and the angular momentum is $n-j$. With this choice, the spin-orbit interaction assumes the form
\begin{eqnarray}\label{epaps:hso}
\hat H_{SO}&&=
\frac{i \hbar \alpha}{\ell}\sum_{j,n}\sqrt{2 j} 
(\hat a^\dag_{j-1,n,\downarrow}\hat a_{j,n,\uparrow}-
\hat a^\dag_{j,n,\uparrow}\hat a_{j-1,n,\downarrow}) \nonumber \\
&&+\frac{ \hbar \beta}{\ell}\sum_{j,n}\sqrt{2 j} 
(\hat a^\dag_{j-1,n,\uparrow}\hat a_{j,n,\downarrow}+
\hat a^\dag_{j,n,\downarrow}\hat a_{j-1,n,\uparrow})~,\qquad
\end{eqnarray}
where the spin quantization axis is along $+\hat z$. 

\subsection{Definition of $f_{1,2}$}

The second-order contribution to the energy from the spin-orbit interaction is
$\delta E=\sum_\alpha |\langle \Omega |\hat H_{SO} |\alpha\rangle|^2/(E_\Omega-E_\alpha)$ where
$|\Omega \rangle$ and $|\alpha \rangle$ denote the ground state and excited states respectively. 
These we suppose are full eigenstates of the interacting Hamiltonian, including LL mixing, 
with energy $E_{\Omega,\alpha}$ and spin quantized along $\vec n$.

This energy correction $\delta E$ corresponds to Eq.~(2) in the main text (when $g=0$). 
The general angular dependence can be obtained applying to Eq.~(\ref{epaps:hso}) a spin rotation to the $\vec n$ direction 
\begin{eqnarray}
&&\hat a^\dag_{j, n, \uparrow}=
(\cos{\theta/2}~\hat a^\dag_{j, n, +}+\sin{\theta/2}~\hat a^\dag_{j ,n, -})e^{i\varphi/2}~,\\
&&\hat a^\dag_{j, n, \downarrow}=
(\sin{\theta/2}~\hat a^\dag_{j, n,+}-\cos{\theta/2}~\hat a^\dag_{j, n, -})e^{-i\varphi/2}~.\quad
\end{eqnarray}
By a straightforward calculation we find that the angular-dependent contribution has the form 
$C_1 m (\alpha^2-\beta^2)\cos\theta+C_2 (\alpha^2+\beta^2+2\alpha\beta \sin2\varphi)\sin^2\theta$.
This can be expressed as in Eq.~(2) in the main text, where $f_{1,2}$ are defined by the following 
identities 
\begin{eqnarray}\label{epaps:f1_def_original}
\eta f_1(\eta,\nu)=&&2J+1
+\frac{1}{p N}\sum_{\alpha,j,n,\mu}\frac{\hbar^2/m\ell^2}{E_\Omega-E_\alpha}\,\mu j \\
&&\times |\langle \Omega|\hat a^\dag_{j-1,n, \mu}\hat a_{j,n, -\mu} 
+\hat  a^\dag_{j,n,-\mu}\hat a_{j-1,n,\mu} |\alpha\rangle|^2~,\nonumber\\
\label{epaps:f2_def_original}
\eta f_2(\eta,\nu)=&&-\frac{1}{p N}\sum_{\alpha,j,n,\mu}\frac{\hbar^2/m\ell^2}{E_\Omega-E_\alpha}\,\mu j \\
&&\qquad\times |\langle \Omega|\hat a^\dag_{j-1,n, +}\hat a_{j,n, \mu}
-\hat  a^\dag_{j,n,-}\hat a_{j-1,n,-\mu}\nonumber \\
&&\qquad-\hat  a^\dag_{j-1,n,-}\hat a_{j,n,-\mu}
+\hat  a^\dag_{j,n,+}\hat a_{j-1,n,\mu} |\alpha\rangle|^2~,\nonumber
\end{eqnarray}
where $2J<\nu<2J+1$ and $p=(\nu-2J)/\nu$ is the ground-state polarization. The right-hand sides
of Eqs.~(\ref{epaps:f1_def_original}, \ref{epaps:f2_def_original}) are difficult to evaluate in general,
but they are found to vanish in the noninteracting case. Therefore, $f_{1,2}$ 
express the contribution to $C_{1,2}$ due to the Coulomb interaction.

\subsection{Calculation of $f_{1,2}$ to leading order}

By expanding the the eigenstates $|\Omega\rangle$, $|\alpha\rangle$ and the corresponding energies
to first order in $\hat V_{ee}$, the following standard third-order contribution is obtained
\begin{eqnarray}\label{epaps:de}
\delta E_3 =
&&\sum_{a,b} \frac{\langle 0 |\hat V_{ee}| a\rangle 
\langle a |\hat H_{SO}|b \rangle \langle b|\hat H_{SO}|0 \rangle + {\rm c.c.}}{(E_0-E_a)(E_0-E_b)}
\nonumber\\
&&+\sum_{a,b} \frac{\langle 0 |\hat H_{SO}| a\rangle 
\langle a |\hat V_{ee}|b \rangle \langle b|\hat H_{SO}|0 \rangle}{(E_0-E_a)(E_0-E_b)}
\nonumber\\
&&-\langle 0 |\hat V_{ee}|0 \rangle \sum_a \frac{|\langle 0 |\hat H_{SO}| a\rangle|^2}{(E_0-E_a)^2}~,
\end{eqnarray}
where $|0\rangle$ is the ground state, that we assume has a fully polarized highest LL, 
and $|a\rangle$, $|b\rangle$ are excited states. All the unperturbed states
are now non-interacting eigenstates, but are chosen to diagonalize $V_{ee}$ to lowest order. 

It is seen from (\ref{epaps:hso}) that the spin-orbit 
interaction produces single $\pm \hbar \omega_c$ excitations, and at the same time changes the angular 
momentum by $\mp 1$. Therefore, the total angular momentum cannot be conserved in the first term of 
(\ref{epaps:de}), which is vanishing. Furthermore, $E_a-E_0=E_b-E_0=\hbar\omega_c$ and
\begin{eqnarray}
\delta E_3 =
\frac{\langle 0 |\hat H_{SO} \hat V_{ee} \hat H_{SO}|0 \rangle
-\langle 0 |\hat V_{ee}|0 \rangle  \langle 0 |\hat H_{SO}^2| 0\rangle}
{(\hbar \omega_c)^2}~,\qquad
\end{eqnarray}
which involves averages of strings of $a^\dag _{j,m,\sigma}$, $a_{j,m,\sigma}$ operators on our 
particular choice of ground state $|0\rangle$. By spin rotation and appropriate evaluation of such averages, 
one obtains an expression in which the only non-trivial matrix elements are of the form
$\langle 0 | \hat a^\dag_{J, m, +} \hat a^\dag_{J, p, +} \hat a_{J, q, +} \hat a_{J, n, +} |0\rangle$.
In fact, also terms containing a string of six $J$ operators appear, but in this case the form of 
$\hat H_{SO}$ is such that a factor $\sum_ n \hat a^\dag_{J, n, \mu} \hat a_{J, n, \mu}$ can be extracted 
to act directly on $|0\rangle$.

Therefore, the final result can be expressed in terms of the following interaction coefficients
\begin{eqnarray}\label{epaps:V_def}
\left. V\right._{i,j}^{h,k}= \frac12 && \sum_{n,m,p,q} 
\langle 0 | \hat a^\dag_{J, m, +} \hat a^\dag_{J, p, +}  
\hat a_{J, q, +} \hat a_{J, n, +} |0\rangle \\
&&\times
\langle \varphi_{i,m}({\bf r}_1)\varphi_{h,p}({\bf r}_2)|
V_{ee}(r_{12})
|\varphi_{k,q}({\bf r}_2)\varphi_{j,n}({\bf r}_1)\rangle ~, \nonumber
\end{eqnarray}
where $V_{ee}(r_{12})=\frac{e^2}{\epsilon r_{12}}-\frac{1}{L^2}\int \frac{e^2}{\epsilon r} {\rm d}{\bf r}$,
which accounts of the the neutralizing background. In the range $0<\nu<1$ we have
\begin{equation}\label{epaps:E3result_1stLL}
\delta E_3 =
\frac{2m \gamma_-^2}{\hbar \omega_c} 
\Big(\left. V\right._{0,0}^{0,0}-\left. V\right._{1,1}^{0,0}
\Big)\cos\theta
-\frac{m \gamma_+^2}{\hbar \omega_c} 
\left. V\right._{1,0}^{0,1} \cos^2\theta ~,
\end{equation}
where we have used $\sin{2\varphi}=\sgn(\alpha \beta)$. Eq.~(\ref{epaps:E3result_1stLL}) is
further simplified using the identity $\left. V\right._{1,0}^{0,1}=\left. V\right._{0,0}^{0,0}-\left. V\right._{1,1}^{0,0}$. For $2<\nu<3$ we have
\begin{eqnarray}\label{epaps:E3result_2ndLL}
\delta E_3 =
&&\frac{2m \gamma_-^2}{\hbar \omega_c} 
\Big(\left. V\right._{1,1}^{0,0}
+ \left. V\right._{1,1}^{1,1}-
2\left. V\right._{2,2}^{1,1}-\frac14\sqrt{\frac{\pi}{2}}\frac{e^2}{\epsilon \ell} \, p N \Big)\cos\theta \nonumber\\
-&&\frac{m \gamma_+^2}{\hbar \omega_c} 
\Big(\left. V\right._{1,0}^{0,1}
+2\left. V\right._{2,1}^{1,2}-2\sqrt{2}\left. V\right._{1,2}^{1,0} \Big) \cos^2\theta 
~.
\end{eqnarray}

Comparing $\delta E=pN[-g\mu_B B/2+(2J+1)\gamma_-^2]\cos\theta+\delta E_3$ with Eq.~(2) in the main text, we obtain explicit formulas for $f_1$, $f_2$ in terms of the $\left. V\right._{i,j}^{h,k}$ coefficients. At $0<\nu<1$
\begin{equation}
f_1(0,\nu)=f_2(0,\nu)=-2 \frac{\left. V\right._{1,0}^{0,1}}{ N e^2/\epsilon \ell}~,
\end{equation}
and at $2<\nu<3$
\begin{eqnarray}
f_1(0,\nu)=\sqrt{\frac{\pi}{8}}-2\frac{\left. V\right._{1,1}^{0,0}
+ \left. V\right._{1,1}^{1,1}-2\left. V\right._{2,2}^{1,1}}{p N ~ e^2/\epsilon \ell}~,\\
f_2(0,\nu)=-2\frac{\left. V\right._{1,0}^{0,1}
+2\left. V\right._{2,1}^{1,2}-2\sqrt{2}\left. V\right._{1,2}^{1,0} }{p N ~ e^2/\epsilon \ell}~.
\end{eqnarray}




\subsection{Calculation of $\left. V\right._{i,j}^{h,k}$}

It is straightforward to obtain $\left. V\right._{i,j}^{h,k}$ 
from the corresponding `generalized' pair-correlation functions 
$\left. g\right._{i,j}^{h,k}$
\begin{equation}\label{epaps:V_fromg}
\left. V\right._{i,j}^{h,k}= \frac{(p\rho)^2}{2} \int \frac{e^2}{\epsilon r_{12}}
[\left. g\right._{i,j}^{h,k}({\bf r}_1,{\bf r}_2)-\delta_{i,j}\delta_{h,k}] {\rm d}{\bf r}_1 {\rm d}{\bf r}_2 ~, 
\end{equation}
where
\begin{eqnarray}\label{epaps:gdef}
\left. g\right._{i,j}^{h,k}({\bf r}_1,{\bf r}_2)= \frac{1}{(p\rho)^2} \sum_{n,m,p,q}
\langle 0 | \hat a^\dag_{J,m,+} \hat a^\dag_{J,p,+}  
\hat a_{J,q,+} \hat a_{J,n,+} |0\rangle \nonumber\\
\times \varphi^*_{i,m}({\bf r}_1)\varphi^*_{h,p}({\bf r}_2)
\varphi_{k,q}({\bf r}_2)\varphi_{j,n}({\bf r}_1) 
~,~\qquad
\end{eqnarray}
and $p\rho=\frac{\nu-2J}{2\pi \ell^2}$ is the fraction of the 
electron density in the highest LL.

The lowest-order case $\left. g\right._{0,0}^{0,0}$ 
is the ordinary pair-correlation function for the ground-state, expressed as a wavefunction
in the lowest LL, as it is customary also for $J>0$ \cite{epaps:JainCambridge07}. 
For a homogeneous isotropic state this is conveniently expressed in the form \cite{epaps:girvin84}
\begin{equation}
\left.g\right._{0,0}^{0,0}= 
1-e^{-|z|^2/2\ell^2}+{\sum_m}^{\prime}
\frac{2}{m!}\left( \frac{|z|^2}{4\ell^2}\right)^m c_m e^{-|z|^2/4\ell^2}~,
\end{equation}
where $z=z_1-z_2$ ($z_{\alpha}=x_{\alpha}+i y_{\alpha}$) and the prime 
indicates a summation over positive odd values of $m$ only. Note that here the 
parameters $c_m$ depend explicitly on the specific ground state $|0\rangle$. At higher order 
analytic expressions are obtained from multiple derivatives of $\left.g\right._{0,0}^{0,0}$,
according to the definitions (\ref{epaps:gdef}) and (\ref{epaps:psidef2}). In particular, the following
recursive relations hold
\begin{eqnarray}
\label{epaps:recursive_rel1}
-i\sqrt{2}\ell\frac{\partial \left.g\right._{i,j}^{h,k}}{\partial z_1}&=& 
\sqrt{j+1}\,\left.g\right._{i,j+1}^{h,k}-\sqrt{i}\,\left.g\right._{i-1,j}^{h,k} ~,\\
\label{epaps:recursive_rel2}
i\sqrt{2}\ell\frac{\partial \left.g\right._{i,j}^{h,k}}{\partial z_1^*}&=&
\sqrt{i+1}\,\left.g\right._{i+1,j}^{h,k}-\sqrt{j}\,\left.g\right._{i,j-1}^{h,k} ~,\\
\label{epaps:recursive_rel3}
-i\sqrt{2}\ell\frac{\partial \left.g\right._{i,j}^{h,k}}{\partial z_2}&=&
\sqrt{k+1}\,\left.g\right._{i,j}^{h,k+1}-\sqrt{h}\,\left.g\right._{i,j}^{h-1,k} ~,\\
\label{epaps:recursive_rel4}
i\sqrt{2}\ell\frac{\partial \left.g\right._{i,j}^{h,k}}{\partial z_2^*}&=&
\sqrt{h+1}\,\left.g\right._{i,j}^{h+1,k}-\sqrt{k}\,\left.g\right._{i,j}^{h,k-1} ~.
\end{eqnarray}
As an example, the expression required to obtain $f_{1,2}$ at $0<\nu<1$ reads
\begin{eqnarray}
\left.g\right._{1,0}^{0,1}=\left(\frac{|z|^2}{2\ell^2}-1\right) e^{-|z|^2/2\ell^2}
-
{\sum_m}^{\prime}
\frac{1}{m!}
\left( \frac{|z|^2}{4\ell^2}\right)^{m-1} \qquad   \\
\times\left[m^2 - (2m+1) \frac{|z|^2}{4 \ell^2}+ \frac{|z|^4}{16 \ell^4}\right]c_m e^{-|z|^2/4\ell^2} ~.
\nonumber
\end{eqnarray}

The explicit form of $\left. V\right._{i,j}^{h,k}$ easily follows. 
In particular, the quantities $\left. V\right._{J,J}^{J,J}$ are required for the total energy of 
a trial wavefunction in a generic LL. For $J=0$ the total energy is $\left. V\right._{0,0}^{0,0}$
\begin{equation}\label{epaps:energyexpansion1}
\frac{\left. V\right._{0,0}^{0,0}}{L^2 (e^2/\epsilon \ell)}
=\frac{\nu^2}{2\pi \ell^2} \bigg[-\sqrt{\frac{\pi}{8}} + 
{\sum_m}^{\prime} \frac{c_m \Gamma(m+1/2)}{m!}\bigg]~,
\end{equation}
while for $J=1$ the energy is $\left. V\right._{1,1}^{1,1}-N\sqrt{\frac{\pi}{8}}\frac{e^2}{\epsilon \ell}$,
where the constant is due to the presence of a filled lowest LL. We obtain
\begin{eqnarray}\label{epaps:energyexpansion2}
\frac{\left. V\right._{1,1}^{1,1}}{L^2 (e^2/\epsilon \ell)}
=\frac{(\nu-2)^2}{2\pi \ell^2}  \bigg[-\frac34 \sqrt{\frac{\pi}{8}} 
+  {\sum_m}^{\prime} \frac{c_m \Gamma(m-3/2)}{64 m!} \nonumber\\
\times (8 m-11) (8 m-3) \bigg]~. \qquad
\end{eqnarray}

Similar expressions for $f_{1,2}$ are also derived and presented in the main text.
For convenience, we repeat them here. At $0<\nu<1$,
\begin{equation}\label{epaps:f12_LLL_epaps}
f_1(0,\nu)=f_2(0,\nu)=\frac{\nu}{2}\sqrt{\frac{\pi}{2}} +
\nu {\sum_m}^\prime \frac{c_m  \Gamma(m-1/2)}{4 m!} ~,
\end{equation}
and at $2<\nu<3$,
\begin{eqnarray}\label{epaps:f1_epaps}
f_1(0,\nu)=\frac{3\nu-2}{8}\sqrt{\frac{\pi}{2}} +
(\nu-2) {\sum_m}^\prime \frac{c_m  \Gamma(m-5/2)}{256 m!} \nonumber\\
\times 3 (8m-15)(8m-5)~,\qquad
\end{eqnarray}
\begin{eqnarray}\label{epaps:f2_epaps}
f_2(0,\nu)=\frac{7(\nu-2)}{8}\sqrt{\frac{\pi}{2}}
+(\nu-2){\sum_m}^\prime \frac{c_m \Gamma(m-5/2)}{256 m!}\nonumber\\
\times (105-112m+64m^2) ~.\qquad
\end{eqnarray}

These can be compared to the energy expansions. 
In particular, (\ref{epaps:f12_LLL_epaps}) to (\ref{epaps:energyexpansion1}) and (\ref{epaps:f1_epaps}, \ref{epaps:f2_epaps}) 
to (\ref{epaps:energyexpansion2}). We do not find any general relation among them, except in the 
simple situation when a single low-order $c_m$ coefficient gives the main contribution. 
Then, $f_{1,2}$ and the energy  are related through the ratios of the prefactors of this particular $c_m$. 
For example, concerning the cusps in $f_{1,2}$, we find that they are generally downward as for the energy, 
with the only exception of $f_1$ in the range $2<\nu<3$. In fact, the ratio of the $c_1$ 
prefactors in (\ref{epaps:f1_epaps}) and (\ref{epaps:energyexpansion2}) is negative, 
which gives an upward cusp for $f_1$.

We also note that the method described in this section is immediately applicable to a 
modified two-body interaction, which for example occurs for a finite thickness of the sample. 
In fact, the expansion of the $\left. V\right._{i,j}^{h,k}$ in terms of the $\{c_m \}$ 
parameterization is obtained by a (possibly numeric) term-by-term integration of the exact expressions 
of the corresponding $\left. g\right._{i,j}^{h,k}$. 

\subsection{Electron-hole symmetry}

For a ground state with $2J<\nu<2J+1$ and fully polarized highest LL, one can 
construct the ground state at $4J+1-\nu$ and same magnetic field $B$ by making 
use of the electron-hole symmetry in the polarized LL. This allows to obtain 
a relation between the $\left. V\right._{i,j}^{h,k}$ at conjugated filling factors. 
At $0<\nu<1$ it reads
\begin{equation}
\left. V\right._{i,j}^{h,k}(1-\nu)=(1-2\nu)\left. V\right._{i,j}^{h,k}(1)+(-1)^{i+j+h+k} \left. V\right._{i,j}^{h,k}(\nu)~,
\end{equation}
and at $2<\nu<3$
\begin{equation}
\left. V\right._{i,j}^{h,k}(5-\nu)=(5-2\nu)\left. V\right._{i,j}^{h,k}(3)+(-1)^{i+j+h+k} \left. V\right._{i,j}^{h,k}(\nu)~.
\end{equation}
The corresponding relations for $f_{1,2}(\eta,\nu)$, are also obtained (if $\eta=0$). 
In the range of $0<\nu<1$ the two coefficients $f_{1,2}$ are equal and
\begin{equation}
f_{1,2}(0,5-\nu)=\frac{(1-2\nu)f_{1,2}(0,1)+\nu f_{1,2}(0,\nu)}{1-\nu}~,
\end{equation}
where $f_{1,2}(0,1)=\sqrt{\frac{\pi}{8}}$. At $2<\nu<3$ we obtain for the function 
$f_2(0,\nu)$ the following relation
\begin{equation}\label{epaps:f2_ehole}
f_{2}(0,5-\nu)=\frac{(5-2\nu)f_{2}(0,3)+(\nu-2) f_{2}(0,\nu)}{3-\nu}~,
\end{equation}
which is not satisfied by $f_1(0,\nu)$. Instead, one has to apply the transformation (\ref{epaps:f2_ehole})
to the function $\tilde f_1(0,\nu)=f_1(0,\nu)-\sqrt{\frac{\pi}{8}}$. Finally, we have $f_{1}(0,3)=
f_{2}(0,3)=\frac78\sqrt{\frac{\pi}{2}}$.

\subsection{Non-analytic features}

\begin{figure}
\includegraphics[width=0.4\textwidth]{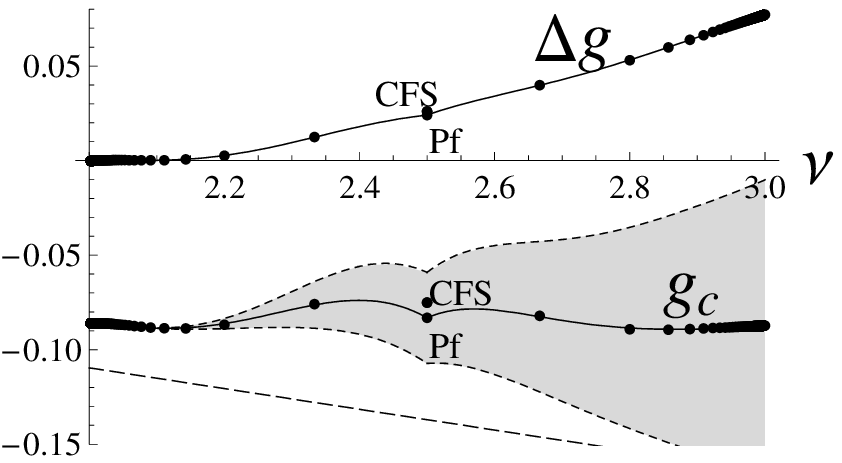}
\caption{\label{epaps:gc_vs_nu_epaps} Values of $g_c$ and $\Delta g$, in the range $2<\nu<3$ (material parameters and $\{c_m\}$ aas in the main text). At $\nu=5/2$ both Composite Fermi Sea (CFS) and Pfaffian (Pf) results are displayed. The solid curves are a guide for the eye, highlighting the expected cusps at the Pfaffian  values. The transition region is given by the gray area and the noninteracting $g_c$ is the lowest dashed line.}
\end{figure}

We comment here about the possibility of a non-analytic dependence of the $f_{1,2}$ coefficients on the filling factor, which is reflected on the dependence of $g_c$ and $\Delta g$. By making use of appropriate trial wave-function we obtained  in the main text the $\{c_m\}$ parameters and therefore discrete values of $g_c$, $\Delta g$ for points in the intervals $0<\nu<1$ and $2<\nu<3$. We reproduce here Fig.~2 of the main text, with the results for the second LL. As done in \cite{epaps:levesque84} for the energy, we smoothly interpolate between the discrete values (see dots and solid lines in Fig.~\ref{epaps:gc_vs_nu_epaps}) and thus extend our results to general $\nu$. 

Although a smooth interpolation might result in a good approximation of the true curve, the presence of cusps is expected at the incompressible states, based on the following theoretical argument. We first notice the similarity of the expansions (\ref{epaps:energyexpansion1},\ref{epaps:energyexpansion2}) of the energy and (\ref{epaps:f12_LLL_epaps}-\ref{epaps:f2_epaps}) for $f_{1,2}$. The presence of cusps in the energy is well known, due to the excitation gaps at the incompressible quantum Hall states \cite{epaps:halperin84}. We conclude that the $\{ c_m \}$ have a non-analytic behavior and that similar cusps in $f_{1,2}$ are expected, due to the $c_m$ dependence of (\ref{epaps:f12_LLL_epaps}-\ref{epaps:f2_epaps}).

While the energy cusps are downward, the same does not hold in general for $f_{1,2}$. Nevertheless, for the particular case $\nu=5/2$ we can infer the qualitative form from our previous results. If at $\nu=5/2$ a noninteracting CF sea were realized, the $g_c$, $\Delta g$ curves would go smoothly through the CFS values shown in Fig.~\ref{epaps:gc_vs_nu_epaps}. As discussed in the main text, this condition might be observed when the temperature exceeds the pairing energy \cite{epaps:willett02}, while at lower temperature an incompressible state is formed. Therefore, also the smooth CFS curves would evolve with temperature to form cusps at $\nu=5/2$. Since the Pfaffian values of $g_c$, $\Delta g$ are lower than the CF sea ones, this suggests the presence of two \emph{downward} cusps in $g_c$ and $\Delta g$ of which the former is more pronounced. 




\end{document}